\documentclass[%
 aip,
 apl,%
 amsmath,amssymb,
 reprint,%
final
]{revtex4-1}

\usepackage{graphicx}

\begin{document}

\preprint{AIP/123-QED}

\title[BaFe$_{1.8}Co_{0.2}$As$_2$ thin film hybrid Josephson junctions]{BaFe$_{1.8}$Co$_{0.2}$As$_2$ thin film hybrid Josephson junctions}

\author{S. Schmidt}
\author{S. D\"oring}
\author{F. Schmidl}%
\author{V. Grosse}
\author{P. Seidel}
\email{paul.seidel@uni-jena.de.}
 \affiliation{Friedrich-Schiller-Universit\"at Jena, Institut f\"ur Festk\"orperphysik, Helmholtzweg 5, 07743 Jena, Germany}%
\author{K. Iida}
\author{F. Kurth}
\author{S. Haindl}
\author{I. M\"onch}

\author{B. Holzapfel}
\affiliation{IFW Dresden, P. O. Box 270116, 01171 Dresden, Germany}%

\date{\today}

\begin{abstract}
Josephson junctions with iron pnictides open the way for fundamental experiments on superconductivity in these materials and their application in superconducting devices. Here, we present hybrid Josephson junctions with a BaFe$_{1.8}$Co$_{0.2}$As$_2$ thin film electrode, an Au barrier and a PbIn counter electrode. The junctions show resistively shunted junction-like current-voltage characteristics up to the critical temperature of the counter electrode of about 7.2\,K. The temperature dependence of the critical current shows nearly linear behavior near $T_\mathrm{C}$. Well-pronounced Shapiro steps are observed at microwave frequencies of 10$-$18\,GHz. Assuming an excess current of 200$\,\mathrm{\mu A}$ at 4.2\,K the effective $I_\mathrm{C}R_\mathrm{N}$ product calculates to 7.9$\,\mathrm{\mu V}$.
\end{abstract}



\maketitle

Examining pnictide-based Josephson junctions is an effective way to understand the nature of superconductivity, since fundamental properties, such as energy gap and order parameter symmetry, can be derived. If the symmetry of pairing differs from conventional s-wave, the behavior of these junctions will change. There exist some theoretical works comparing different types of symmetry and the resulting junction properties. Some new experiments with Josephson junctions were proposed \cite{Parker2009,Tsai2009,Ota2009,Inotani2009,Chen2010,Wu2009}.

The first observations of Josephson effects in the pnictides were reported for doped BaFe$_2$As$_2$ (Ba-122) by  Zhang et al. \cite{Zhang2009}, who fabricated hybrid Josephson junctions with a conventional s-wave counter-electrode (lead) and Ba$_{1-x}$K$_x$Fe$_2$As$_2$ single crystals ($T_\mathrm{C}$ about 20\,K) in \textit{c}-direction. The Pb electrode was used in two geometries, point contact tip and planar thin film of PbIn, respectively. They reported rather conventional Josephson behavior. Zhou et al. \cite{Zhou2009} realized a hybrid corner junction with a BaFe$_{1.8}$Co$_{0.2}$As$_2$ single crystal ($T_\mathrm{C}=22\,\mathrm{K}$) and a Pb electrode. The Fraunhofer-like pattern for the critical Josephson current in an external magnetic field seems to exclude d-wave symmetry. But as shown theoretically \cite{Parker2009}, this simple corner junction geometry can not provide exact information about the symmetry for extended s-wave. The first thin film Josephson junctions with Ba-122 were realized on (La,Sr)(Al,Ta)O$_3$ bicrystals by Katase et al. \cite{Katase2010}, who observed RSJ-like $I$-$V$ characteristics up to 17\,K, but an $I_\mathrm{C}R_\mathrm{N}$ product of only $60\,\mathrm{\mu V}$ at 4.2\,K. This kind of junctions was recently used for a first d.c. superconducting quantum interference device\cite{Katase2010a}. 

In this letter we present the fabrication of hybrid Ba-122/Au/PbIn thin film Josephson junctions and their electrical properties. Josephson effects were observed up to the critical temperature of the PbIn counter-electrode.

\begin{figure}
\centering
		\includegraphics[width=0.9\columnwidth]{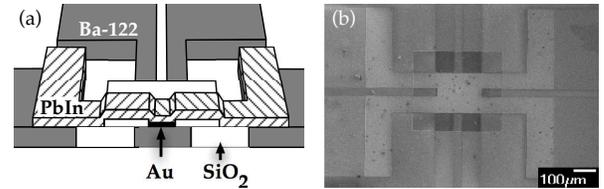}
	\caption{(a) Schematic illustration based on the photolithographic mask design. (b) Overview of a prepared junction structure acquired by scanning electron microscopy, SEM.}
	\label{fig:tunnel_scheme}
\end{figure}
Co-doped Ba-122 thin films have been used as pnictide electrodes, which were prepared by a standard pulsed laser deposition technique. Details on film preparation can be found elsewhere\cite{Iida2010}. The films show a very good surface quality with a RMS roughness of less than $1\,$nm enabling the deposition of a complete Au layer as barrier material with a thickness from 5 to 10\,nm. No significant change in the RMS roughness occurred during this deposition process. Covering the whole Ba-122 thin film with gold also prevents contamination during future photoresist processes.

The gold covered iron pnictide base electrodes were patterned via ion beam etching, IBE, (500\,eV beam voltage, 10$^{-3}$\,Acm$^{-2}$ ion beam density). Our junction design (Fig.~\ref{fig:tunnel_scheme}a) precisely allows the determination of $T_\mathrm{C}$ and $I_\mathrm{C}$ of both base and counter electrode, as well as for the whole
Josephson junction in four-point geometry. The junction windows with areas from $3\times3\,\mathrm{\mu m^2}$ to $100\times100\,\mathrm{\mu m^2}$ were defined by sputtered SiO$_2$ (thickness of 100\,nm) in a lift-off process. Prior to the SiO$_2$ deposition we removed the Au shortcut in the junction area via IBE. The counter electrode was produced in another lift-off process via thermal evaporation of PbIn with a thickness of 150\,nm. Au wires (diameter of $25 \,\mathrm{\mu m}$) were used as bonding material.

\begin{figure}
\centering
	\includegraphics[width=0.9\columnwidth]{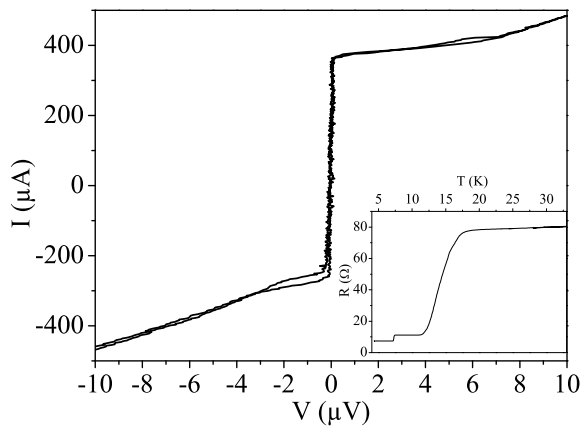}
	\caption{$I$-$V$ characteristics of the Josephson junction measured bidirectional in current bias mode. An hysteretic difference between increasing and decreasing bias currents is visible in the negative branch. The inset shows the resistance versus temperature of a $30\times30\,\mathrm{\mu m^2}$ junction measured with a bias current of $10\,\mathrm{\mu A}$.}
	\label{fig:I-U_R-T}
\end{figure}

The results we present here are measured in four-point geometry at samples with a junction area of $30\times30\,\mathrm{\mu m^2}$ and a gold barrier thickness of 5\,nm. The resistance of the junction shows two drops at temperatures of 7.2\,K and 15\,K, respectively (inset of Fig.~\ref{fig:I-U_R-T}). By comparing these values with the critical temperatures of our electrode materials Ba-122 \cite{Iida2010} and PbIn measured prior to the preparation, we can assure that the fabrication processes has no influence on both $T_\mathrm{C}$ values.

The $I$-$V$ characteristics show hysteretic behavior (Fig.~\ref{fig:I-U_R-T}). It seems that there are multiple branches and sometimes the junction jumps between them. This behavior leads to noticeable variations of the critical current, $I_\mathrm{C}$, between different curves measured at constant temperature, especially at lower temperatures. There is  also a clear asymmetry between the positive and the negative critical current, which may be caused by trapped flux.

\begin{figure}
	\centering
		\includegraphics[width=0.9\columnwidth]{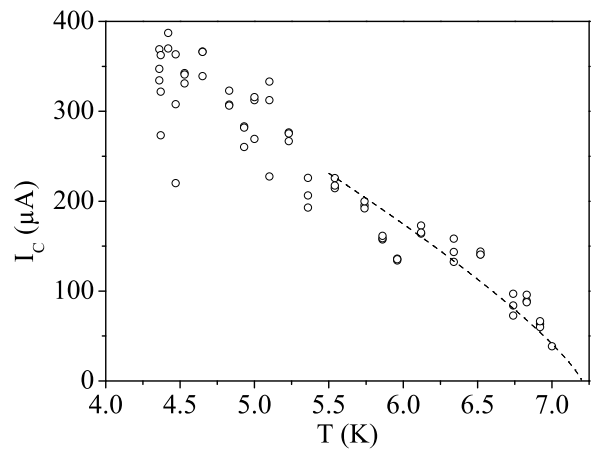}
		\caption{Critical Josephson current versus temperature, symbols are the experimental data while the dotted line is a guide for the eye.}
	\label{fig:Ic-T}
	
\end{figure}

Thus we determined a critical current from several $I$-$V$ characteristics at constant temperatures. The temperature dependence of $I_\mathrm{C}$ is shown in Fig.~\ref{fig:Ic-T}. At 4.2\,K it is about 350$\,\mathrm{\mu}$A, decreases nearly linear with the tempera\-ture and is zero at 7.2\,K, which corresponds to the $T_\mathrm{C}$ of the PbIn counter electrode. The normal state resistance, $R_\mathrm{N}$ of the junction is $53\,\mathrm{m\Omega}$. So we obtain a formal $I_\mathrm{C}R_\mathrm{N}$ product of $18.4\,\mathrm{\mu V}$. This is nearly the same magnitude determined by other groups \cite{Zhang2009a,Katase2010}. But extrapolating the linear part of the $I$-$V$ characteristic (Fig.~\ref{fig:I-U_R-T}) to zero voltage gives rise to assume an excess current, $I_\mathrm{ex}$, of about $200\,\mathrm{\mu}$A, lowering the effective $I_\mathrm{C}$ to 150$\,\mathrm{\mu}$A and the effective $I_\mathrm{C}R_\mathrm{N}$ product to $7.9\,\mathrm{\mu V}$. For positive driving currents the $I$-$V$ characteristic converges the linear dependence at $7\,\mathrm{\mu V}$, which corresponds roughly to the effective $I_\mathrm{C}R_\mathrm{N}$ value. From the junction area size, the critical Josephson current density, $J_\mathrm{C}$, may be calculated to $39\,\mathrm{Acm^{-2}}$, while the sheet resistance $\rho_\mathrm{N}$ of the junction is $5800\,\Omega cm^{-2}$. Even if it could be expected from the small $I_\mathrm{C}R_\mathrm{N}$ product, because of the low value of $J_\mathrm{C}$ and the high value of $\rho_\mathrm{N}$, it can be ruled out that the Josephson junction is formed by a grain boundary. We rather assume the junction to be of the SNS, SINS or SINIS type, with N as normal, I as insulating and S as superconducting layer.

\begin{figure}
	\centering
		\includegraphics[width=0.9\columnwidth]{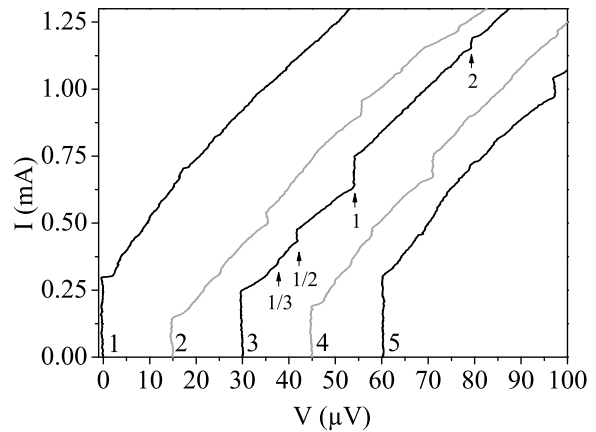}
	\caption{$I$-$V$ characteristic under microwave irradiation at different frequencies.
	(1) without microwave irradiation. (2) 10\,GHz. (3) 12\,GHz. (4) 13\,GHz. (5) 18\,GHz. Each curve is shifted by $15\,\mathrm{\mu V}$ to the one before. Shapiro steps at the 12\,GHz curve are marked by arrows. The index of each step denotes multiples of $2eV/hf$.}	
	\label{fig:I-V}
\end{figure}

\begin{figure}
\includegraphics[width=0.9\columnwidth]{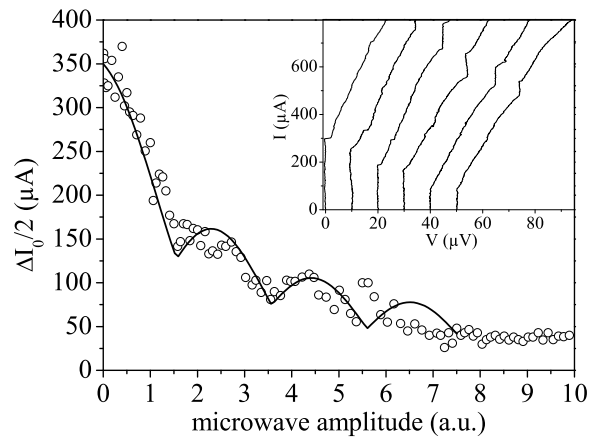}
\caption{\label{fig:stepheight}Critical current versus microwave voltage at 12\,GHz. The straight line denotes a Bessel simulation with an underlying exponential decay shown by the dotted line. This exponential behavior is probably thermally induced. The fit parameters are $\beta_\mathrm{C}=2.5$, $\Omega = 2.85$ and $R_\mathrm{N} = 53\,\mathrm{\mu\Omega}$ with an excess current $I_\mathrm{ex}=200\,\mathrm{\mu A}$ at zero microwave amplitude. The inset shows some $I$-$V$ characteristics at different microwave voltages. Each curve is shifted by $10\,\mathrm{\mu V}$ to the one before.}
\end{figure}

Under microwave irradiation Shapiro steps were observed (Fig.~\ref{fig:I-V}). Depending on the microwave frequency and power, steps up to the fifth order occur. Also subharmonic steps can be observed matching to step indices of $1/2$ and $1/3$. The power dependence of the zeroth Shapiro step is shown in Fig.~\ref{fig:stepheight}. To model this power dependence we applied the resistively and capacitively shunted junction approach like in ref.~\onlinecite{Seidel1991} for which the following differential equation has to be solved: 
\begin{equation}\label{eqn:rcsj}
\beta_\mathrm{C}\ddot{\varphi}+\dot{\varphi}+\sin\varphi=i_\mathrm{b}+i_\mathrm{m}\sin\Omega\tau\ .
\end{equation}
Here, $\beta_\mathrm{C}=2eI_\mathrm{C}R_\mathrm{N}C/\hbar$ is the McCumber parameter and $\tau=\omega_0 t$ is the normalized time with $\omega_0=2eI_\mathrm{C}R_\mathrm{N}/\hbar$ the characteristic frequency of the junction. The second term on the right side of eq.~(\ref{eqn:rcsj}) accounts for the microwave irradiation with the microwave amplitude, $i_\mathrm{m}$, the normalized microwave frequency,  $\Omega=\omega_\mathrm{m}/\omega_0$, and the dc bias current, $i_\mathrm{b}$. For high frequencies or amplitudes the solution of eq.~(\ref{eqn:rcsj}) can be approximated to the well known Bessel behavior given by:
\begin{equation}\label{eqn:bessel}
\Delta I_k=2I_\mathrm{C}\left|J_k(A)\right|
\end{equation}
with $A=\frac{i_\mathrm{m}}{\Omega}\frac{1}{\sqrt{1+\beta_\mathrm{C}^2\Omega^2}}$ and $J_k$ as the Bessel function.

From our experimental data we find that even at high microwave amplitudes the critical current is not completely suppressed, see Fig.~\ref{fig:stepheight}. This justifies the introduction of an excess current, which additionally appears to depend on microwave power. We obtain a good agreement with eq.~(\ref{eqn:bessel}) by assuming an exponential decrease of $I_\mathrm{ex}$ with microwave amplitude. For these calculations we used $\Omega=2.85$ and $\beta_\mathrm{C}=2.5$, which was determined from the average $I$-$V$ hysteresis without microwave irradiation. The exponential fit of $I_\mathrm{ex}$ gives a value of $200\,\mathrm{\mu A}$ at zero microwave power which agrees well with the value determined from the $I$-$V$ characteristics.
As discussed above the switching between different branches often gives an even more pronounced hysteresis in the $I$-$V$ characteristics than shown in Fig.~\ref{fig:I-U_R-T}. In those measurements the corresponding $\beta_\mathrm{C}$ can reach values of up to 10 indicating an unusually high parallel capacitance for a SNS-type Josephson junction. This seeming discrepancy, however, is rather related to the junction layout than to the junction type. High stray capacitances arising from the insulation layers and the substrate can be expected.
As can be shown by numerical simulation of $I$-$V$ characteristics such high values of $\beta_\mathrm{C}$ and a thermal noise contribution can lead to a lowering of the maxima compared to the Bessel behavior or even a complete suppression of parts of the Shapiro steps \cite{Seidel1991, Busse}. Additionally, non-homogeneous junction properties and a non-linear power dependent excess can result in deviations from the ideal behavior.

In summary, we fabricated planar thin film BaFe$_{2}$As$_{2}$/Au/PbIn hybrid Josephson
junctions and investigated their electrical behavior. The $I$-$V$ characteristics
are resistively shunted junction-like with  a non-linear excess current and a hysteresis corresponding
to a McCumber parameter $\beta_\mathrm{C}$ of about 2 to 3. The $I_\mathrm{C}R_\mathrm{N}$ product is in the order of
$10\,\mathrm{\mu V}$ at 4.2\,K. The temperature dependence of the critical Josephson current close to $T_\mathrm{C}$ is nearly linear as expected for SNS junctions. Well pronounced Shapiro steps
were observed for frequencies of 10 to 18\,GHz. Their power dependence can be
described by the conventional Bessel behavior assuming a non-linear and power
dependent excess current. The results will be used to further develop Josephson junctions based on pnictide thin films. This would also give a chance to investigate the intrinsic Josephson effects in La-1111 just reported for single crystals\cite{Kashiwaya}. Additionally, the realization of Josephson junctions along the $ab$-plane is in progress.\\


This work was partially supported by DFG within SPP 1458 (projects SE 664/15-1 and HA 5934/3-1). We thank L. Schultz, M. Kidszun and A. Grib sincerely for their contributions.


%

\end{document}